\documentclass[aps,prb,twocolumn,superscriptaddress,10pt]{revtex4-1}
\usepackage[utf8]{inputenc}
\usepackage{libertine}
\usepackage[libertine]{newtxmath}
\usepackage{mathtools,graphicx,microtype,bm}
\usepackage[cal=boondoxo,frak=boondox]{mathalfa}

\begin{document}
\title{Effective Mass of Bound Electron Pairs in Two-Dimensional Materials with a Gapped Band Spectrum}
\author{Vladimir A.\ Sablikov}
\email[E-mail:]{sablikov@gmail.com}
\author{Bagun~S. Shchamkhalova}
\email[E-mail:]{s.bagun@gmail.com}
\affiliation{Kotelnikov Institute of Radio Engineering and Electronics, Fryazino Branch, Russian Academy of Sciences, Fryazino, Moscow District, 141190, Russia}

\begin{abstract}
Bound electron pairs formed due to the peculiarities of the band dispersion of electrons in crystals attract much interest because they can carry charge and spin even in the absence of band conductivity. However, such an important parameter of bound pairs as the effective mass is still poorly understood. We carry out this study for materials described by the Bernevig-Hughes-Zhang model in the electron-hole symmetric case and find a clear relationship between the effective mass and the energy of a bound pair at rest. The dependence of mass on energy has a specific form for each of different types of pairs, but a common feature is the change of the mass sign when energy passes through the middle of the band gap. The sign is negative when the energy is in the lower half of the gap, and positive in the upper half.
\end{abstract}

\maketitle

\textit{Introduction.} The pairing of electrons due to the Coulomb interaction attracts a long-standing interest, not only because of problems of high-temperature superconductivity~\cite{kagan2013modern,RevModPhys.62.113}, but in recent years also in connection with the general interest to the electron-electron (e-e) interaction effects in materials with non-trivial band states~\cite{Rachel_2018, Kotov}. The greatest attention is paid to the pairing mechanism caused by the peculiarities of the band dispersion in a wide variety of materials and low-dimensional systems with both a gapless and gapped spectrum. Due to the specific structure of basis atomic orbitals and their hybridization, various types of bound electron pairs (BEPs) of this nature can form in such systems. They were studied for topological insulators (TIs)~\cite{Sablikov}, Dirac semimetals~\cite{Portnoi}, graphene and bigraphene~\cite{Sabio2010,Lee2012,MahmoodianEntinEPL2013,MarnhamShytovPRB2015,Egger}, carbon nanotubes~\cite{Hartmann}, systems with flat band~\cite{Torma,Iskin}, systems of cold atoms in optical traps~\cite{winkler2006repulsively,hou2017many}, chiral cold atomic systems~\cite{tai2017microscopy}. Even in nanophotonic systems with a periodic structure, two-photon quasiparticles can form~\cite{PhysRevA.95.053866}. The pairing mechanism is often interpreted as a result of the formation of a negative reduced mass~\cite{Gross}, although in some cases such an idea seems too oversimplified, since the equation of the relative motion of pairing electrons generally has a more complex form than the usual motion of a particle in the e-e interaction potential~\cite{Sablikov}.

The BEPs formed in this way are unusual charge and spin carriers, which can play an important role especially in the cases where their energy is in the band gap and under non-equilibrium conditions. Interest in them is stimulated by the recently established fact that BEPs are sufficiently resistant to radiative decay~\cite{https://doi.org/10.1002/pssr.201900358,https://doi.org/10.1002/pssb.202000299}. Therefore, it is now important to study the transfer of BEPs under the action of external fields and, first of all, to find their effective mass. Until now, this issue has been extremely poorly studied.

The energy of a BEP was studied as a function of its momentum for several very simplified forms of the crystal potential~\cite{MahajanJPhysA2006,Hai_2014,Hai_2018}. However this approach does not allow to draw definite conclusions regarding the magnitude and even sign of the effective mass and, moreover, the mechanism of its formation. In addition it does not reflect the structure of atomic orbitals forming electronic states which are important for currently relevant low-dimensional materials.

The transport of bound pairs of repulsive particles has been more consistently studied in relation to cold atomic systems within the framework of the Bose- and Fermi-Hubbard models, which take into account both on-site and nearest neighbor repulsion, but are limited to considering only one lower band of continuum states. This approach allows one to find out different families of the bound pairs, depending on the model parameters of the interaction~\cite{Valiente_2009,PhysRevA.79.032108,HU2021127634} and even study the dynamics of bound-pair packets induced by the scattering with a single particle~\cite{PhysRevA.79.032108,PhysRevA.83.052102}. Of most interest in view of the subject of the present study is the fact that the effective mass of bound pairs depends on the e-e interaction parameters used in the model~\cite{HU2021127634,doi:10.7566/JPSJ.90.104702}. However, no fundamental conclusions have been drawn about the dependencies of the effective mass on the interaction potential and the single particle effective mass in the band of continuum states.

The present paper is aimed at studying the effective mass of BEPs for a wide class of currently relevant two-dimensional materials with a two-band spectrum, in which BEPs are formed by mixing of atomic orbitals of both electron-like and hole-like basis states. Accordingly, it can be expected that the effective mass of such BEPs is largely determined by the composition of the atomic orbitals that form them. This is an important factor that has not yet been studied and, as will be shown, radically changes the magnitude and even the sign of the effective mass. In this situation, the calculation of the effective mass of BEPs is a very nontrivial problem, since the composition of atomic orbitals is determined by the equations of motion of the particles for specific interaction potential and the momentum of the pair~\cite{Sablikov}. Therefore, the composition of the atomic orbitals of a moving BEP varies with its momentum and can be very different from that of a BEP at rest. It is also interesting that the motion of the center of mass of the pair is not separated from the relative motion of the electrons, and therefore the motion of the pair as a whole leads to a change in the binding energy of the BEP.

We develop an approach to solving this problem and calculate the effective mass within the Bernevig-Hughes-Zhang (BHZ) model~\cite{BHZ} which describes a wide class of two-dimensional materials with strong spin-orbit interaction in both topological and trivial phases. As a result, we have found that there is a general relationship between the effective mass of a BEP and its energy at rest. In the case of the symmetric BHZ model, the effective mass of BEPs is positive when the energy is above the middle of the gap of the two-particle band spectrum, and negative when the energy is in the lower half of the gap. Near the middle of the gap, the effective mass vanishes for all types of BEPs.

\textit{Model approaches.} We will study BEPs within the framework of the BHZ~\cite{BHZ} model, which qualitatively well describes many materials and is well suited for studying the effects of e-e interaction when the interaction radius is large enough, as in systems with a low electron density. The model is based on the $\mathbf{k\cdot p}$ approximation. The single-particle states are presented in the four-band basis $(|E\uparrow\rangle,|H\uparrow\rangle,|E\downarrow\rangle,|H\downarrow\rangle)^T$, composed of electron and hole orbitals with spin up and down. For simplicity, we restrict ourselves to the case when the model has electron-hole symmetry. 

Since the spin projection $S_z$, is conserved the two-electron states can be categorized into two types by their spin~\cite{Sablikov}:

(i) singlets with $S_z=0$ formed by two-particle basis orbitals $(|E\uparrow,E\downarrow\rangle,|E\uparrow,H\downarrow\rangle,|H\uparrow,E\downarrow\rangle,|H\uparrow,H\downarrow\rangle)^T$ and a similar state with permuted particles,

(ii) two triplets with $S_z=\pm 1$ formed by orbitals $(|E\uparrow,E\uparrow\rangle,|E\uparrow,H\uparrow\rangle,|H\uparrow,E\uparrow\rangle,|H\uparrow,H\uparrow\rangle)^T$ and $(|E\downarrow,E\downarrow\rangle,|E\downarrow,H\downarrow\rangle,|H\downarrow,E\downarrow\rangle,|H\downarrow,H\downarrow\rangle)^T$.

To be specific, we will consider the singlets in topological phase. For triplets, the results are quite similar. 

Properties of the BEPs essentially depend on the composition of the orbitals forming them, since it is the contributions of the electron and hole orbitals that are decisive factor determining the effective mass of BEPs. In its turn this composition is self-consistently determined by the equation of motion and depends on the interaction potential. As a result, the BEPs formed by electrons, both of which occupy predominantly hole orbitals, are very different from BEPs in which one electron predominantly occupies a hole orbital while the other occupies the electron one. We classify these states as BEPs of the first and second groups, respectively.

All possible types of the BEPs with zero total momentum were studied previously~\cite{Sablikov,https://doi.org/10.1002/pssr.201900358,https://doi.org/10.1002/pssb.202000299}. Here we extend this study to the case of finite but small total momentum $\mathbf{K}$. For this purpose, we represent the BEP wave function in the form:
\begin{equation}
  \Psi_S(\mathbf{R},\mathbf{r})=\Psi_{\mathbf{K}}(\mathbf{r})e^{i\mathbf{KR}},
\end{equation}
where $\mathbf{R}$ and $\mathbf{r}$ are the coordinates of the center of mass and the relative position of the electrons, and use the Hamiltonian derived in Ref.~\cite{Sablikov} and given there by Eq.~(11), for singlet BEPs. We represent this Hamiltonian in a more convenient form by separating the part describing the motion of a BEP as a whole
\begin{equation}\label{HS}
  H=H_0+H_K\,,
\end{equation}
where $H_0$ is the Hamiltonian of the BEP at rest
\begin{equation}
  \hat{H}_0(\mathbf{\hat{k}})=
\begin{pmatrix}
  2(\mathbf{\hat{k}}^2-1+v(\mathbf{r})) & a\mathbf{\hat{k}}_{-}&a\mathbf{\hat{k}}_{+}&0\\
 a\mathbf{\hat{k}}_{+} & 2v(\mathbf{r})& 0 &a\mathbf{\hat{k}}_+\\
 a\mathbf{\hat{k}}_{-}& 0 &2v(\mathbf{r}) & a\mathbf{\hat{k}}_{-}\\
 0&a\mathbf{\hat{k}}_{-}&a\mathbf{\hat{k}}_{+}&2(-\mathbf{\hat{k}}^2+1+v(\mathbf{r}))\\
\end{pmatrix},
\label{H0}
\end{equation}
and $H_K$ is the $\mathbf{K}$-dependent component,
\begin{equation}
  \hat{H}_K(\mathbf{\hat{k}},\mathbf{K})=
\begin{pmatrix}
 K^2/2& -a\mathbf{K_{+}}/2&a\mathbf{K_{+}}/2&0\\
 -a\mathbf{K_{+}}/2 &2\mathbf{\mathbf{K}\!\cdot\!\hat{k}}&0&a\mathbf{K_{+}}/2\\
 a\mathbf{K_{-}}/2&0&-2\mathbf{\mathbf{K}\!\cdot\!\hat{k}} & -a\mathbf{K_{-}}/2\\
 0&a\mathbf{K_{-}}/2&-a\mathbf{K_{+}}/2 &-K^2/2\\
\end{pmatrix}.
\label{HK}
\end{equation}
Here and in what follows it is convenient to use dimensionless notations. All values of the energy dimension are normalized to $|M|$. The distances are normalized to $\sqrt{|B/M|}$ where $B$ and $M$ are the parameters of the BHZ model, $B$ describes the dispersion in the electron and hole bands, $M$ is the mass term. $v(r)$ is the normalized e-e interaction potential per electron. The parameter $a=A(B M)^{-1/2}$ is important since it describes the hybridization of the electron and hole bands, where the parameter $A$ describes the mixing the electron and hole bands in the BHZ model~\cite{BHZ}. $\mathbf{\hat{k}}$ is the relative momentum operator, $\hat{k}_{\pm}=\hat{k}_{x}\pm i \hat{k}_{y}$, and $\mathbf{K_{\pm}}=K_x\pm i K_y$.

Wave functions of the singlet BEPs are four-rank spinors in the two-particle basis:
\begin{equation}
  \Psi_K(\mathbf{r})=(\psi_3,\psi_4,\psi_7,\psi_8)^T,
\end{equation} 
and the spinor components are defined from the Schr\"odinger equation $(H_0+H_K)\Psi_K= E \Psi_K$, where $E$ is the two-particle energy. 

For simplicity, we restrict ourselves to considering situations where the band dispersion has only one central extremum in the conduction and valence bands. In the case of the topological phase $BM>0$, this happens when $a\ge \sqrt{2}$. At $a<\sqrt{2}$, the band dispersion has a shape of the mexican hat and the study of electronic states is more complicated. In the case of the trivial phase, the band dispersion has only one extremum in each band at any $a$. But in what follows we will keep in mind the topological phase. 

First conclusions about the effective masses can be made directly from the Hamiltonian~(\ref{HS}) in the case of small interaction potential, $v(r)\ll 1$, for BEPs of the first group with the energy level $E$ close to the top of the valence band of the two-particle spectrum, $(E +2)\ll 2$. Simple analysis shows that in the spinor $\Psi_K$, the component $\psi_3$ predominates, but the presence of other components reflecting the hybridization of the electron and hole orbitals, is very important. As a result, the following equation is obtained for $\psi_3$ for $K\ll 1$:
\begin{equation}
 \left[\left(1-\frac{a^2}{2}\right)\left(\hat{k}^2+\frac{K^2}{4}\right) + v(r) \right]\psi_3=\left(\frac{E}{2}+1\right)\psi_3\,. 
\end{equation}
It is seen that the energy has the form: $E=-2+2\varepsilon_n +K^2/(2M^*)$, where $\varepsilon_n$ is the binding energy of the BEP per electron at rest with respect to the valence band, $M^*$ is the effective mass of the BEP which is equal to
\begin{equation}
  M^*=\frac{2}{2-a^2}\,.
\label{M*}
\end{equation}
Obviously, $M^*$ is exactly equal to twice the effective mass near the central extremum in the valence band, $m^*=1/(2-a^2)$. 

The relative motion with quantized spectrum $\varepsilon_n$ is described by the reduced effective mass $m_r$ which equals half the band mass, $m_r=m^*/2$. When $a>\sqrt{2}$, the reduced effective mass is negative and therefore the repulsive e-e interaction leads to the formation of BEPs. However, in this case the effective mass $M^*$ of the BEPs is also negative for any shape of $v(r)$. This means that the BEPs are unstable, since Coulomb forces acting between the pairs lead to their mutual attraction and formation of many-electron complexes with energy levels in the gap. This obviously reflects the well-known process of renormalization of the band gap due to e-e interaction~\cite{Inkson_1976}.  

The situation, however, changes with an increase in the e-e interaction potential, since the interaction amplitude significantly affects the composition of the orbitals that form the BEPs~\cite{Sablikov,https://doi.org/10.1002/pssr.201900358}. For example, at a low interaction potential, BEPs of the first group are formed mainly by electrons of the valence band. An increase in the interaction amplitude leads to additional mixing of the orbitals of the conduction band, which leads to a change in both the reduced effective mass and the effective mass of the BEPs, but these masses obviously change in different ways. Below we study how the effective mass of a pair changes with the e-e interaction potential in the case of a small center-of-mass momentum, $K\ll 1$, using perturbation theory.

\textit{A perturbative approach.} When the center-of-mass momentum is small enough, the problem can be solved by treating the Hamiltonian $H_K$ in Eq.~(\ref{HS}) as a perturbation. Such an approach was successfully used to study excitons, especially in materials with complicated valence band spectrum~\cite{DRESSELHAUS195614,PhysRevB.11.3850}. Expansion of the energy of a pair in momentum in the second order gives the effective mass of the pair. The perturbation series can be constructed on the basis of two-particle states of the Hamiltonian~(\ref{H0}). Two-particle eigenstates of the Hamiltonian~(\ref{H0}) were studied in Refs.~\cite{Sablikov,https://doi.org/10.1002/pssr.201900358,https://doi.org/10.1002/pssb.202000299} in the case where the parameter $a\ge \sqrt{2}$. The total spectrum of two-particle states contains both bound states with a discrete spectrum in the band gap and unbound states in the bands. Below we briefly present these states.

There are two groups of the bound states that differ in their spatial structure and atomic orbital composition. In polar coordinates ($r,\varphi)$, the bound states are described by the following spinors: 
\begin{equation}\label{Psi_m}
  \Psi_m(\mathbf{r,\varphi})\!=\!
  \begin{pmatrix}
  \psi_{3m}(r) \\ i\psi_{4m}(r) e^{i\varphi} \\ i\psi_{7m}(r) e^{-i\varphi} \\ \psi_{8m}(r)
  \end{pmatrix}
  e^{im\varphi}\,,
 \end{equation}
where $m$ is the angular quantum number $m=0, \pm 1, \pm 2, \dots$, $\psi_{3,4,7,8}(r)$ are real functions. The states of the fist group $|m, n_r\rangle$ are additionally characterized by a radial quantum number $n_r$, which appears when the radius of the e-e interaction $r_0$ is large enough. Their energy spectrum $E_{m,n_r}$ and wave functions were studied in Refs.~\cite{Sablikov,https://doi.org/10.1002/pssr.201900358,https://doi.org/10.1002/pssb.202000299} in the case of a step-like potential of e-e interaction $v(r)=v_0\Theta(r_0-r)$. The main feature of the first group states is that their energy tends to the band gap bottom with decreasing the interaction amplitude $v_0$. At $v_0\ll 1$, both electrons forming these pairs occupy predominantly the valence band orbitals. 

The states of the second group, on the contrary, do not exhibit a clearly defined radial quantum number. Nevertheless, at a given angular quantum number $m$, there are two states, which we label with indices $A$ and $B$. The states $|m,A\rangle$ and $|m,B\rangle$ differ in energy for a given $v(r)$, and we will assume that the energy of the $A$ state is lower than that of $B$. As $v_0$ decreases to zero, their energies $E_{m,A}$ and $E_{m,B}$ tend to the middle of the gap. It is important to keep in mind that in the symmetric model, the middle of the band gap of the two-particle spectrum is a specific energy. An infinite set of\textit{ unbound }two-particle states with one electron in conduction band and the other in valence band and zero total momentum of the pair resides at this energy.  The effective mass of a BEP is determined, roughly speaking, by the contributions of the conduction and valence bands to the orbitals that form a bound state. The BEPs of the second group are formed by one electron residing predominantly in the conduction band and the other in the valance band. In the limit $v_0\to 0$, the system is symmetric, and the conduction and valence bands contribute to the bound state equally. Therefore, the total effective mass M* is expected to be zero as the sum of the positive and negative masses of the electrons forming the bound pair. However, at finite $v_0>0$ the symmetry is broken, and the contribution of the conduction band becomes dominant, so that M* is expected to be positive. 

Unbound states of two electrons with zero total momentum $|k, \lambda_1, \lambda_2\rangle$ are characterized by the value of relative wave vector $k$ of electrons which is defined at infinity, $r\to \infty$, where the e-e interaction potential vanishes, and indexes $\lambda_{1,2}$ indicating in which band of the single-particle spectrum each of the two electrons is located~\cite{https://doi.org/10.1002/pssr.201900358,https://doi.org/10.1002/pssb.202000299}. $\lambda$ takes two values $\lambda=1$ and $\lambda=-1$ for the conduction and valence bands respectively. The unbound state energy reads
\begin{equation}
E_{k, \lambda_1, \lambda_2}=(\lambda_1+\lambda_2)\sqrt{(1-k^2)^2+a^2k^2}\,.
\end{equation}  

To be specific, we consider how the motion of the center of mass changes the energy of a BEP of the first group, which at rest was in the state $|m,n_r\rangle$. A perturbation expansion to terms in $K^2$ leads to the following result:
\begin{equation} \label{E(K)}
\begin{split}
E_{m,n_r}(K) & = E_{m,n_r}+\langle m,n_r|H_K|m,n_r\rangle\\
+&\sum_{m'\ne m,n'_r\ne n_r}\frac{\langle m,n_r|H_K|m',n'_r\rangle \langle m',n'_r|H_K|m,n_r\rangle}{E_{m,n_r}-E_{m',n'_r}}\\
+&\sum_{m'\ne m,X=A,B}\frac{\langle m,n_r|H_K|m',X\rangle \langle m',X|H_K|m,n_r\rangle}{E_{m,n_r}-E_{m',X}}\\
+&\sum_{\lambda_1,\lambda_2}\int \frac{d^2k}{4\pi^2}\frac{\langle m,n_r|H_K|k,\lambda_1,\lambda_2\rangle \langle k,\lambda_1,\lambda_2|H_K|m,n_r\rangle}{E_{m,n_r}-E_{k,\lambda_1,\lambda_2}}\,.
\end{split}
\end{equation}  
Here the first string in the right-hand side takes a very simple form:
\begin{equation}
\langle m,n_r|H_K|m,n_r\rangle=\pi \frac{K^2}{2}\int_{0}^{\infty}\!dr r \left(|\psi_{3;m,n_r}|^2-|\psi_{8;m,n_r}|^2\right)
\end{equation}
and describes the change in energy caused by the spinor components corresponding to two configurations in which both particles are in the electron, $\psi_{3;m,n_r}$, and hole, $\psi_{8;m,n_r}$, states with positive and negative masses respectively. The second and third strings describe the effect of electron transitions between different states of the BEP. It is essential that the transitions are possible only when $m$ is changed by $\pm 1$. The fourth string is caused by electron transitions from BEPs into unbound states. It is clear that the contribution of all electron transitions between different states of the pair is proportional to $K^2$. Thus the variation of the BEP energy with $K$ has the form $\Delta E(K) = \alpha K^2$ and is described by an isotropic effective mass $M^*=(2 \alpha)^{-1}$, in dimensionless units.

\textit{The effective mass.} We will now calculate the effective mass $M^*$ for different types of BEPs using Eq.~(\ref{E(K)}). The main goal of this study is to clarify how the effective mass depends on the amplitude and radius of the e-e interaction.

The wave functions required to calculate $M^*$ are found using the model of two-electron states developed in Refs.~\cite{Sablikov,https://doi.org/10.1002/pssr.201900358,https://doi.org/10.1002/pssb.202000299} and some simplifying assumptions. The main simplification is the use of the step-like potential with amplitude $v_0$ and radius $r_0$ that models the short-range interaction of a finite radius. This simplification allows one to solve exactly the two-particle problem within the BHZ model. Another simplification relates to the spectrum of the BEPs. We assume that the energy gap between the discrete levels is large enough, and for this reason the summation over the indexes $n_r$ or $X$ in Eq.~(\ref{E(K)}) can be limited to only one value. This is obviously justified when the interaction radius is small enough. And one more simplification relates to the wave functions of the unbound two-particle states with continuous spectrum. When calculating the matrix elements, we neglect the change in these wave functions due to the e-e interaction. This simplification is justified by our recent study of the matrix elements of the radiative decay of BEPs, which have very similar form. We found that the distortion of the wave functions of unbound states due to the e-e interaction only weakly changes the matrix elements~\cite{https://doi.org/10.1002/pssb.202000299}.

Results of the calculations are as follows.

\begin{figure}
	\centerline{\includegraphics[width=0.95\linewidth]{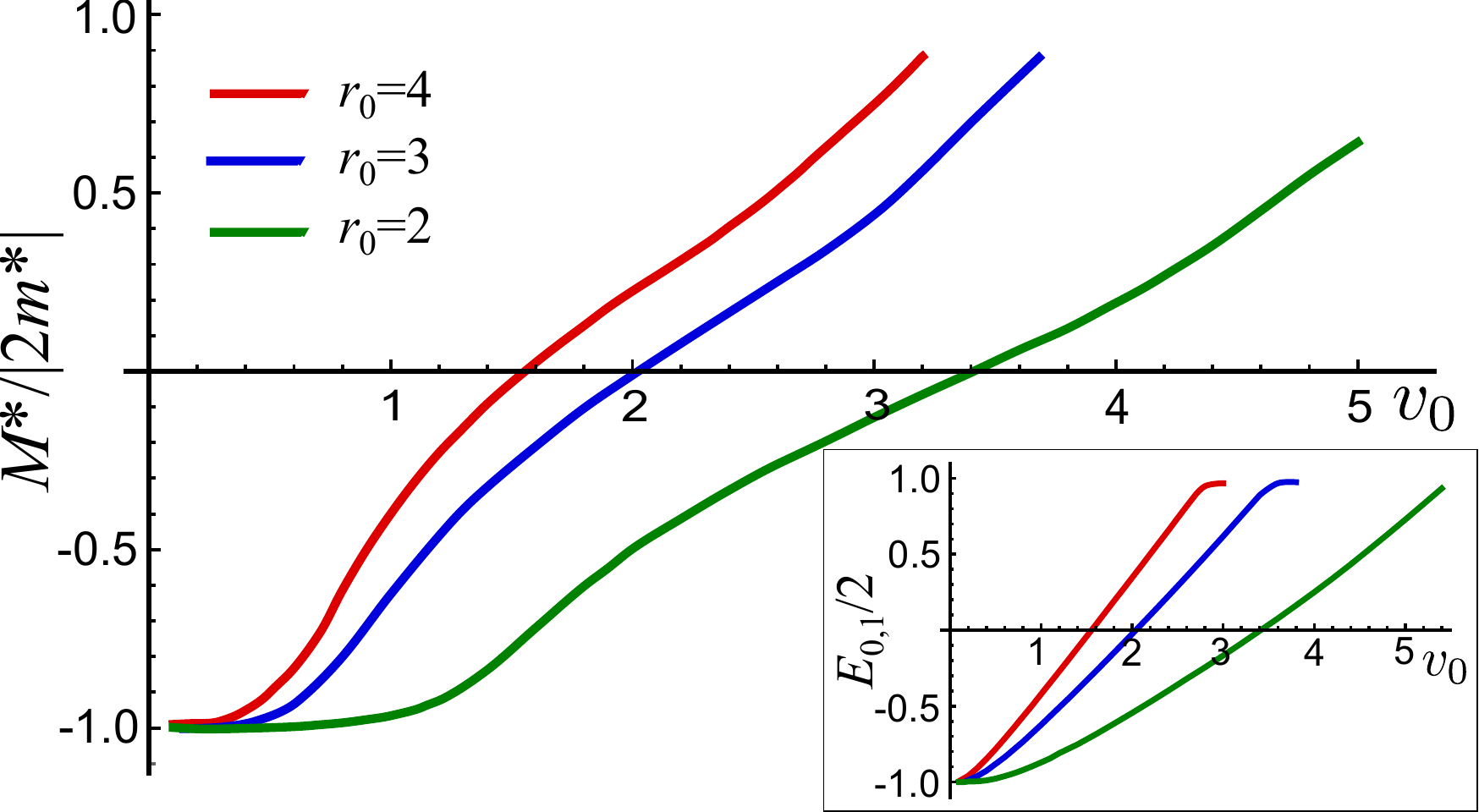}}
	\caption{Dependence of the normalized  effective mass $M^*/(2|m^*|)$ of the BEP of the first group on the interaction potential amplitude $v_0$ for various interaction radius $r_0=2$, $r_0=3$ and $r_0=4$. The insert shows the corresponding energies of the BEPs at rest versus $v_0$. The hybridization parameter used in the calculation is $a=2.1$.}
	\label{fig1}
\end{figure}

First, consider the BEPs of the first group for $a > \sqrt{2}$. The effective mass $M^*$ as a function of the amplitude $v_0$ of the e-e interaction is presented in Fig.~\ref{fig1} for a variety of the interaction radius $r_0$. Here, it is convenient to consider the effective mass $M^*$ normalized to $2|m^*|$. Thus, the normalized effective mass of BEPs with the energy close the valence band, as expected from Eq.~(\ref{M*}), is to be $-1$.

Fig.~\ref{fig1} shows that the normalized effective mass $M^*$ really tends to $-1$ in the limit $v_0\to 0$ and is negative when $v_0$ is not large enough. However, $M^*$ goes to zero and becomes positive when the potential exceeds a critical value which depends on the interaction radius. 

\begin{figure}
	\centerline{\includegraphics[width=0.9\linewidth]{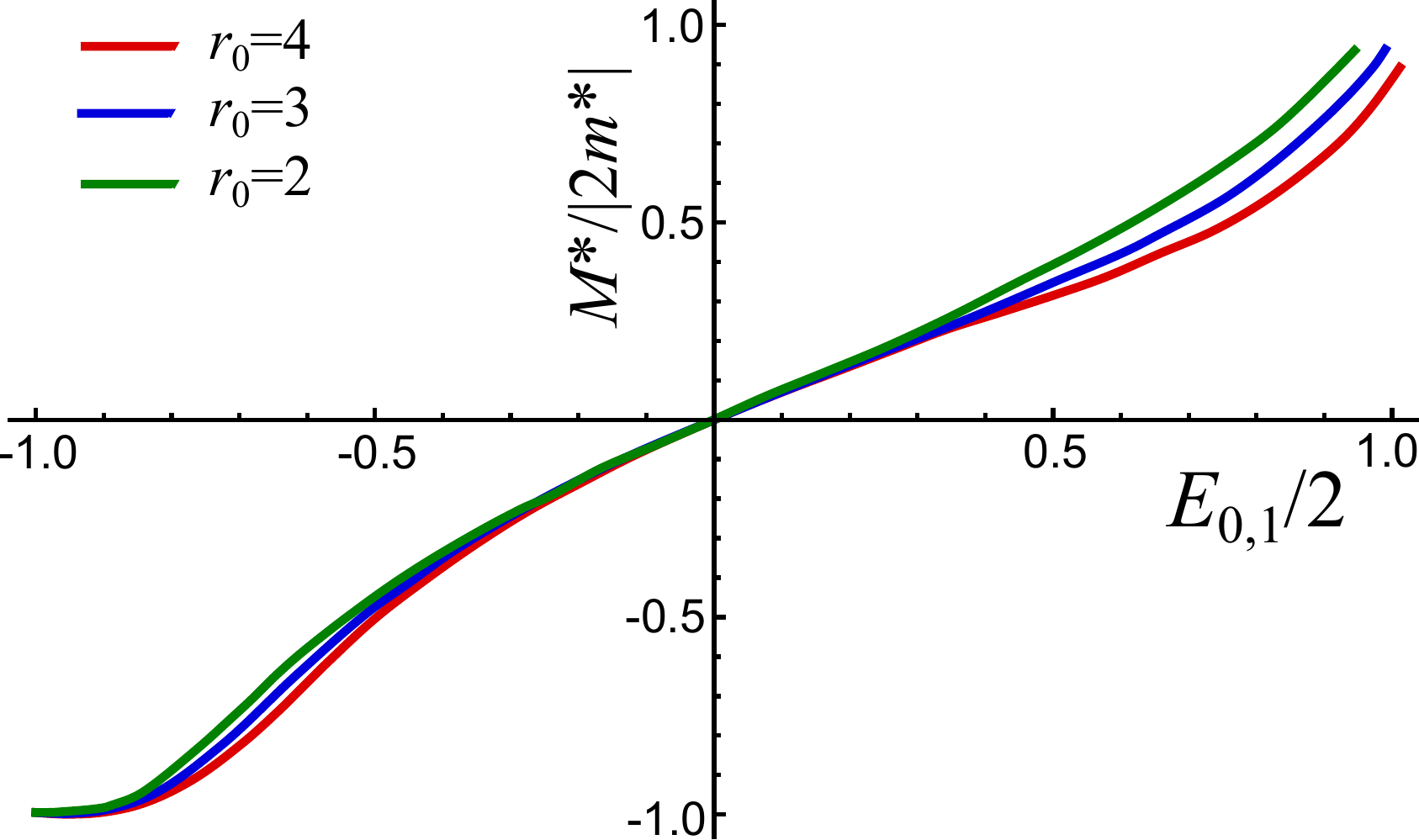}}
	\caption{The effective mass of the BEP of the first group as a function of the bound-state energy at rest for different radii of the e-e interaction $r_0 = 2, 3, 4$.}
  \label{fig2}
\end{figure}

It is interesting to clarify how $M^*$ changes with $v$ on an energy scale relevant to the characteristic energy of BEPs. Within the frame of the perturbation theory we expect that $M^*$ can be expressed in terms of parameters of the unperturbed state. Such an energy parameter is the bound-state energy $E_{m=0, n_r=1}$ at rest. In its turn, this energy is determined by the potential amplitude and radius, as illustrated in the insert in Fig.~\ref{fig1}. Excluding $v_0$ from the functions $M^*(v_0)$ and $E_{0, 1}(v_0)$, we arrive at the mass $M^*$ as a function of $E_{0, 1}$. The result is shown in Fig.~\ref{fig2}. It turns out that this function changes quite slightly with $r_0$. 

In such a way we arrive at an unexpected conclusion that the effective mass of the first group BEPs is close to a universal function of the bound state energy at rest, as shown in Fig.~\ref{fig2}. Here, the deviation of all three lines from the universal one is noticeable only at an energy close to the conduction band. Of course, the conclusion about a slight change in the shape of the dependence of $M^*$ on $E_{0, 1}$ with a change in $r_0$ is limited by simplified assumptions used in these calculations, and is hardly universal for a wide class of functions $v(r)$. Nevertheless, this is a curious fact. 

However, the main conclusion from these results is that the effective mass changes its sign in the middle of the gap and is positive only in the upper half of the band gap. Changing the hybridization parameter $a$ leads to a change in the form of dependence of $M^*$ on $E_{0,1}$, but this main feature remains unchanged for all types of BEPs. Below we illustrate this by considering two cases for completely different types of BEPs.

\begin{figure}
	\centerline{\includegraphics[width=0.9\linewidth]{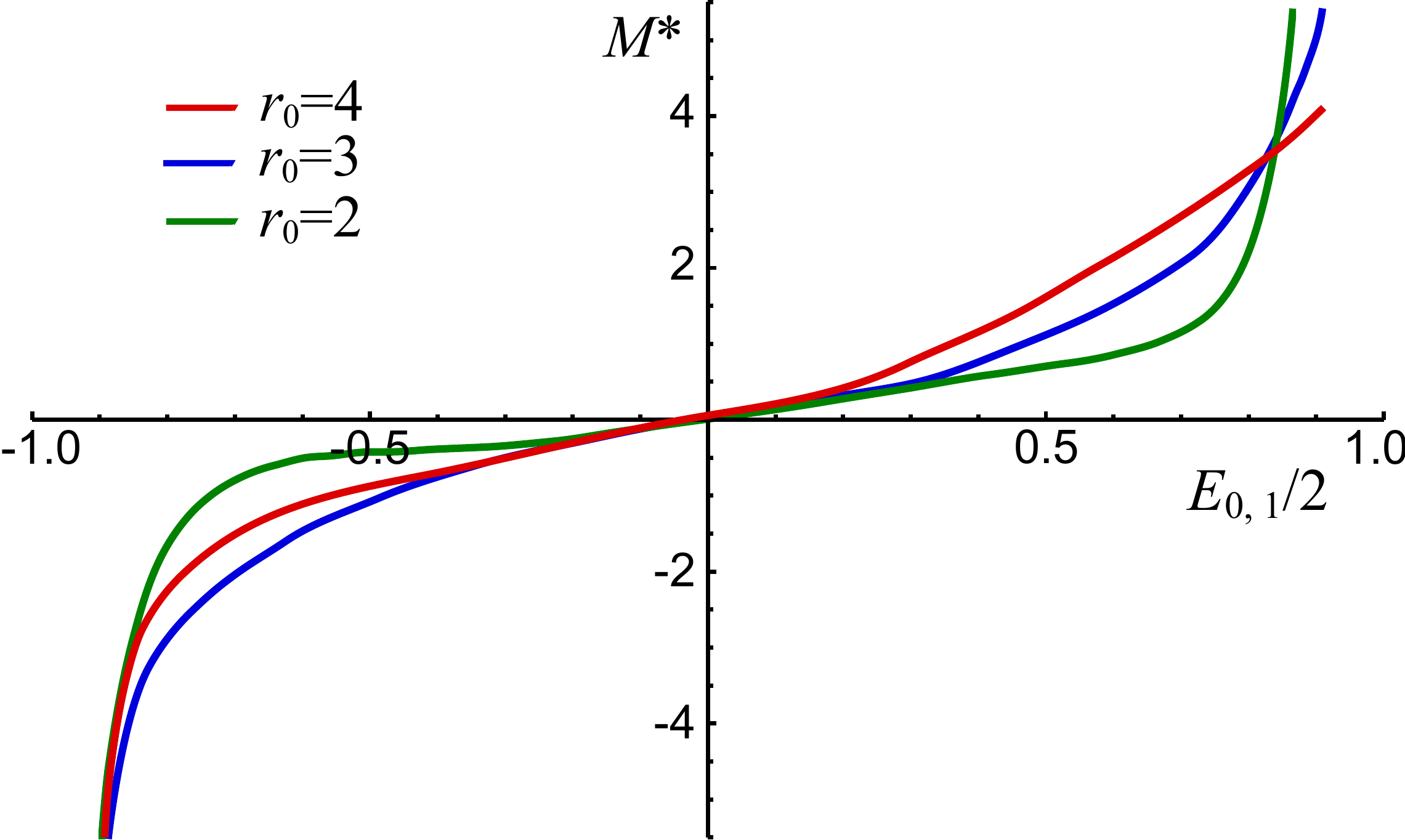}}
	\caption{The effective mass $M^*$ of the BEPs of the first group as a function of the bound-state energy at rest in the case of $a=\sqrt{2}$ for $r_0=2, 3, 4$.}
	\label{fig3}
\end{figure}

Of special interest is the limiting case where $a=\sqrt{2}$. In this case the band dispersion near the valence band extremum becomes almost flat, $\varepsilon+1 \sim k^4$, and the band effective mass $m^*$ formally turns to infinity. Accordingly, the energy of the BEPs of the first group strongly increases. Really the bound-state energy is determined by a reduced effective mass which is formed as a result of hybridization of electron and hole orbitals in a wide range of $k$ and turns out to be finite. In this case, the effective mass of BEPs, $M^*$, as a function of the bound-state energy is shown in Fig.~\ref{fig3} for three values of the e-e interaction radius. It is seen that the above conclusion that $M^*$ turns to zero near the middle of the band gap is also valid in this case, although the shape of the lines describing $M^*$ as a function of $E_{0,1}$ changes. 

\begin{figure}
	\centerline{\includegraphics[width=0.95\linewidth]{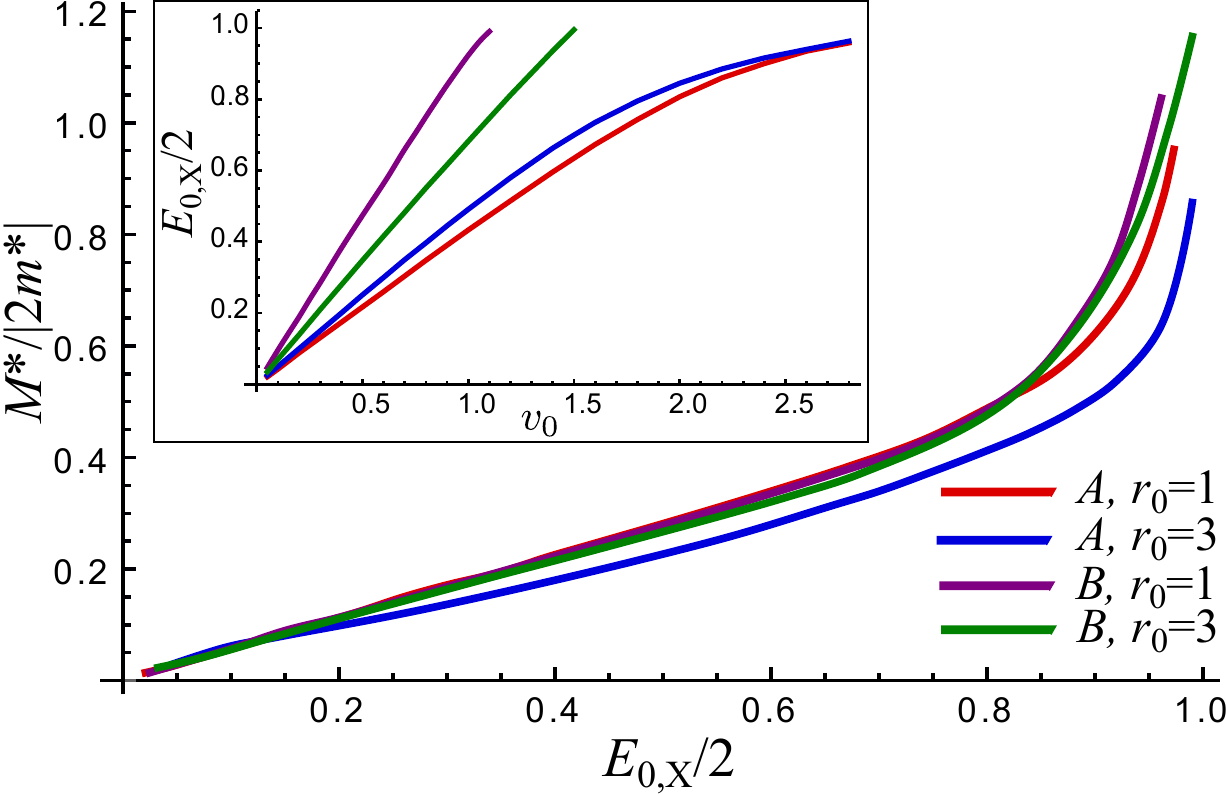}}
	\caption{The effective masses of two BEPs, $|0,A\rangle$ and $|0,B\rangle$, of the second group as functions of their binding energy at rest for two radii of the e-e interaction. The insert shows the bound state energies as functions of the amplitude of e-e interaction potential. The model parameters are $a=2.1$, $r_0=1, 3$.}
	\label{fig4}
\end{figure}

Now we turn to the BEPs of the second group. At $K=0$ they are described by the wave functions $|m, X=A,B\rangle$ and have respectively the energy levels $E_{m,X=A,B}$, both of which lie above the middle of the gap. The peculiarity of these states is that one of the electrons is predominantly in the conduction band, while the other is in the valence band. Since the electron and hole orbitals are almost equally involved in the formation of the BEPs, it can be expected that in the limit of small $v_0$ the effective mass of the pair is close to zero. However, in reality, this electron-hole equality is violated by the interaction potential, and the effective mass becomes finite. We have calculated the mass for both BEPs of the second group indexed above as $A$ and $B$. The energy $E_{m,X}(K)$ of a moving BEP is described by equations similar to Eq.~(\ref{E(K)}), not shown here. In this way we arrive at the results shown in Fig.~\ref{fig4} for both states for one value of the hybridization parameter and two interaction radii.  

It is clearly seen that the effective mass $M^*$ is positive for both states in agreement with our previous conjecture that $M^*$ is positive when the bound state energy lies in the upper half of the gap. In the limit $v_0\to 0$ the effective mass goes to zero, as expected. The dependence of $M^*$ on the bound state energy again is close to an universal one for both states $A$ and $B$, but the deviations from it are much larger than in the case of the first group states.

\textit{Conclusion.} We have studied the effective mass of BEPs that form due to peculiarities of the band dispersion of electronic states composed of electron and hole orbitals in materials with two-band spectrum. Specific calculations were carried out within the frame of the symmetric BHZ model for singlet BEPs in the topological phase, but the main results are of a more general nature. Our calculations show that they are qualitatively valid for both triplet pairs and the trivial phase. The problem of finding the dispersion equation for different types of BEPs  with a finite momentum $\mathbf{K}$ is solved perturbatively by expanding the energy of a moving BEP in $\mathbf{K}$.  

The main conclusion is that the effective mass $M^*$ of the BEPs is defined by their energy $E_{K=0}$ at rest. $M^*$ is negative if $E_{K=0}$ is below the middle of the gap of the two-particle band spectrum and $M^*$ is positive if $E_{K=0}$ lies in the upper half of the band gap. $M^*$ turns to zero when $E_{K=0}$ is near the middle of the gap. Moreover, our results suggest that the form of the function $M^*(E_{K=0})$ is specific for each type of BEPs and, under certain conditions, weakly depends on the parameters of the e-e interaction potential, such as the interaction radius.

The step potential is a good model approximation of the short-range interaction potential, which is realized in realistic systems due to screening processes. This simplification is widely used in the literature, see for example the Ref.~\cite{Sabio2010}. The main its advantage is the possibility to solve the problem exactly, as in the case of bound electron pairs. The results obtained using this simplification are confirmed at the qualitative level for other types of short-range interactions, although specific results may differ quantitatively. In particular, the fact that the effective mass M*of the BEP is negative when its energy is close to the valence band is correct for an arbitrary shape of the potential, see Eq.~(\ref{M*}). With an increase in the bound state energy, the value of the effective mass decreases. 

The fact that $M^*$ is negative in a wide range of energy indicates the instability of the electron system to the formation of many-electron complexes and appearance of many-electron states in the lower part of the forbidden gap. In essence, this reflects the well known process of the gap renormalization due to e-e interaction in many-electron systems~\cite{Inkson_1976}.

Stable BEPs of the first group can form when the e-e interaction is strong so that the bound state energy is large enough. In the symmetric BHZ model, those BEPs are stable whose rest energy is in the upper half of the band gap. This allows one to estimate the condition under which stable BEPs of the first group can appear in the case of the Coulomb interaction. The bound state energy can be roughly estimated by the effective Rydberg $E_R$, which is calculated according to Ref.~\cite{PhysRevA.43.1186} using the band effective mass $m^*$ for the symmetric BHZ model. In this way, we find that the BEP energy is in the upper half of the band gap when $E_R/|M|>1$, and $E_R/|M|\sim e^4[\epsilon^2(a^2-2)|BM|]^{-1}$, with $\epsilon$ being the dielectric constant. This gives a constraint on the material parameters under which the BEPs are stable. A numerical estimate for specific materials shows, for example, that when the material parameters are close to those of HgTe, this constraint can only be met if $a$ is close to $\sqrt{2}$. For materials with a lower dielectric constant, the estimates are more favorable.
 
In contrast, the BEPs of the second group are stable even if the e-e interaction is not so strong. These BEPs are formed by electrons one of which is mainly in the conduction band and the other in the valence band. They are most promising for implementation in materials with strong spin-orbit interaction. This raises the question of a more detailed study of pairs of this type, in particular, in materials with broken particle-hole symmetry, as is the case in many materials used in experiments.

One of the main physical conclusions of our study that the effective mass of a BEP is largely determined by the composition of atomic orbitals can be illustrated by comparing the above results with the results of other theories based on models that do not take into account the mixing of atomic orbitals in the structure of bound pair states. As an example, we will discuss this for the results recently obtained in the framework of the lattice models of BEPs~\cite{winkler2006repulsively,HU2021127634}.

It is obvious that BEPs of the second group are lost in the above tight-binding models, as long as they are limited to consideration of one band of a single-particle spectrum. In this case, the BEPs of the second group are not possible, since they are formed by orbitals of different bands. Therefore we can compare only the BEPs of the first group. At a small amplitude of the interaction potential, the effective mass of these BEPs is negative according to our theory, similarly as according to the Bose–Hubbard model~\cite{winkler2006repulsively}. However, an increase in the interaction potential leads to an increase in the effective mass in the Bose–Hubbard model, while our two-band theory predicts a decrease in the effective mass and even a change in its sign. The reason for this discrepancy is obviously that an increase in the interaction potential in the two band model leads to an increase in the fraction of the conduction band orbitals in the wave function of the BEPs. In a single band model, this is impossible. In addition, the bound states studied in the framework of the BHZ model have the topological properties inherent in this model, which leads to qualitatively new types and properties of pairs in comparison with single-band models.

\textit{Acknowledgments.} This work was carried out in the framework of the state task and supported by the Russian Foundation for Basic Research, project No.~20-02-00126.

\bibliography{BEP_m}
\end{document}